\documentclass[aps,prl,graphicx,twocolumn,a4paper]{revtex4-1}

\usepackage[T1]{fontenc}
\usepackage{graphicx}
\usepackage{dcolumn}
\usepackage{bm}

\begin{document}


\title{Detection and imaging of the oxygen deficiency in single crystalline YBa$_{\text{2}}$Cu$_{\text{3}}$O$_{\text{7}-\delta}$ thin films using a positron beam}

\author{M. Reiner$^1$, T. Gigl$^1$, R.~Jany$^2$, G.~Hammerl$^2$, and C. Hugenschmidt$^1$}

\affiliation{$^1$Lehrstuhl E21 at Physics Department and FRM\,II at Heinz Maier-Leibnitz Zentrum (MLZ), Technische Universit\"at M\"unchen, James-Franck Stra\ss e, 85748 Garching, Germany}
\affiliation{$^2$Experimental Physics VI, Center for Electronic Correlations and Magnetism, University of Augsburg, Universit\"atsstra\ss e 1, 86135 Augsburg, Germany}

\date{\today}

\begin{abstract}
Single crystalline YBa$_{\text{2}}$Cu$_{\text{3}}$O$_{\text{7}-\delta}$ (YBCO) thin films were grown by pulsed laser deposition (PLD) in order to probe the oxygen deficiency $\delta$ using a mono-energetic positron beam. 
The sample set covered a large range of $\delta$ (0.191<$\delta$<0.791) yielding a variation of the critical temperature $T_{\text{c}}$ between 25 and 90\,K.
We found a linear correlation between the Doppler broadening of the positron electron annihilation line and $\delta$ determined by X-ray diffraction (XRD).
Both, the origin of the found correlation and the influence of metallic vacancies, were examined with the aid of ab-initio calculations that allowed us (i) to exclude the presence of Y vacancies and (ii) to ensure that positrons still probe $\delta$ despite the potential presence of Ba or Cu vacancies.
In addition, by scanning with the positron beam the spatial variation of $\delta$ could be analyzed.
It was found to fluctuate with a standard deviation of up to $0.079(5)$ within a single YBCO film.
\end{abstract}

\maketitle

High temperature superconductivity \cite{bednorz} with a maximum $T_{\mathrm{c}}$ of $92\,$K in YBCO \cite{PhysRevLett.58.908,1347-4065-26-4A-L314} is strongly influenced by the oxygen deficiency $\delta$ and the order of oxygen atoms (see e.\,g. refs. \cite{PhysRevB.73.180505,PhysRevB.74.014504,Matic2013189}).
Both, the deeper understanding of the $T_{\mathrm{c}}\left(\delta\right)$ dependence and the precise adjustment of $T_{\mathrm{c}}$ require information about the oxygen vacancies on a microscopic level.
Thin films are extraordinarily suited to gain fundamental insight into the elementary properties of YBCO since the application of PLD using SrTiO$_{\mathrm{3}}$ (STO) substrates enables epitaxial growth of YBCO in single crystalline quality with well defined stoichiometry.
Beyond standard characterization techniques like electrical transport measurements, XRD or electron microscopy, positrons with their unique sensitivity to open-volume defects \cite{RevModPhys.60.701,west,RevModPhys.66.841} reveal valuable additional information.
So far, positron annihilation techniques have been applied in various studies to investigate YBCO bulk materials with positron lifetime spectroscopy \cite{PhysRevB.42.8078,PhysRevB.43.10399,0953-8984-1-23-020,Nieminen19911577}, measurement of the angular correlation \cite{PhysRevLett.60.2198,1402-4896-1989-T29-019} and the Doppler broadening \cite{Smedskjaer198856,PhysRevB.36.8854} of the annihilation radiation.
In particular, positrons were found to be sensitive to the open volume formed by both metallic vacancies and oxygen deficiency in bulk samples of YBCO \cite{RevModPhys.66.841,Nieminen19911577}.

In the present study, we applied depth dependent Doppler Broadening Spectroscopy (DBS) and Coincident DBS (CDBS) with a slow positron beam \cite{RevModPhys.60.701,Coleman}, which allows us to probe films with a thickness of up to several hundreds of nanometers.
DBS is particularly sensitive to the momentum of valence electrons, which has been reported to depend on the oxygen deficiency in YBCO \cite{Smedskjaer198856,PhysRevB.36.8854,PhysRevLett.60.2198,1402-4896-1989-T29-019}. 
For this reason, using a spatially resolving high-intensity positron beam enables both, depth dependent measurements of $\delta$ and imaging of the lateral homogeneity in the YBCO films.
In addition, CDBS allows for elemental specific examination of atoms surrounding the open volume in YBCO. 
In order to emphasize the experimentally gained results, detailed calculations of CDB spectra taking into account metallic vacancies were performed.

YBCO thin films were grown at once by PLD on commercially available STO substrates ($10\times 10$\,mm$^2$) using a KrF laser. 
The substrates were cut into four quadratic pieces, cleaned, heated up to 760\,$^{\circ}$C at a background oxygen pressure of  0.25\,mbar for the thin film deposition and annealed at 400\,mbar.
Three samples (A2, A3, A4) were individually heat treated at a constant temperature of 400\,$^{\circ}$C at different background pressures $p_\mathrm{temp}$ for varying times $t_\mathrm{temp}$ in order to influence their oxygen deficiency (see Tab.\,\ref{tab:samples}).
One thin film sample was left as-grown (A1) as reference.
The quality of the single crystalline YBCO films was routinely checked by XRD, and the c-axis parameter was evaluated by $\Theta$-$2\Theta$-scans for the determination of $\delta$ \cite{Benzi2004625}.
The spatial variation of $\delta$ designated by $\sigma_{\delta}$ was obtained from spatially resolved DBS (see below).
Electrical transport measurements confirmed superconductive behavior for all samples (see $T_c$ in Tab.\,\ref{tab:samples}).
The thickness of a YBCO film grown under identical conditions was determined as 210$\pm$10\,nm by electron microscopy.

\begin{table}[b!]
	\centering
		\begin{tabular}{cccccc}
		\hline
		\hline
		Sample & $\delta$ & $\sigma _{\delta}$ & $T_{\mathrm{c}}$(K)  & $t_{\mathrm{temp}}$(min) & $p_{\mathrm{temp}}$(mbar)\\
		\hline
		A1 & 0.191 & 0.019(1) & 90 & n.\,a. & n.\,a. \\
		A2 & 0.475 & 0.079(5) & 60 & 30 & 2$\cdot$10$^{-2}$ (O$_{\mathrm{2}}$) \\
		A3 & 0.641 & 0.048(3) & 60 & 30 & 10$^{-7}$ \\
		A4 & 0.791 & 0.031(2) & 25 & 50 & 10$^{-7}$ \\
		\hline
		\hline	
		\end{tabular}
	\caption{Parameters of the YBa$_2$Cu$_3$O$_{7-\delta}$ thin films: The oxygen deficiency $\delta$ was determined by XRD; $\sigma _{\delta}$ describes the spatial variation of $\delta$ in the films obtained from spatially resolved DBS (Fig.\,\ref{fig:Sxy}); $T_{\mathrm{c}}$ is the measured critical temperature.}
	\label{tab:samples}
\end{table}

A mono-energetic beam enables depth dependent (C)DBS by implanting positrons with a kinetic energy $E$ up to several keV into the specimen.
The mean penetration depth $\bar{z}$ scales with $E$: $\bar{z}=\frac{A}{\rho}\cdot E^n$; 
$A$ and $n$ are material dependent parameters of the Makhovian implantation profile (A=3.76 $\mu$g/(cm$^2$keV$^n$) and $n=1.64$ for the examined system), and $\rho$ is the mass density. 
After thermalisation within a few picoseconds, the positron diffuses through the sample before it annihilates with an electron after typically 100 - 200\,ps. 
During diffusion the positron can get trapped in attractive potential wells formed by open volume defects where the annihilation probability with core electrons is lower. 
Thus the lower mean momentum of the annihilating electrons with the longitudinal component $p_{\mathrm{L}}^-$ leads to a smaller Doppler shift $\Delta E=\frac{1}{2}\,c\,p_{\mathrm{L}}^-$ of the annihilation $\gamma$-quanta ($c$ is the velocity of light). 
In DBS, $\Delta E$ is measured with high purity Ge detectors (energy resolution 1.4\,keV at 511\,keV).
The Doppler broadening of the annihilation line is commonly evaluated by the lineshape parameter $S$ which is defined by the number of counts in the central region of the 511\,keV photopeak (here $\Delta E < 0.84$\,keV) divided by the total number of counts.
The depth profiles $S(E)$ allow for the extraction of the positron diffusion length $L_+$, which is significantly reduced when positrons are trapped in open volume defects.

In CDBS, a strongly enhanced peak-to-background ratio is achieved by detecting both annihilation $\gamma$-quanta in coincidence \cite{PhysRevLett.77.2097,1998JPCM...1010383M}.
Therefore, high Doppler shifts caused by the annihilation of core electrons can be measured and hence element-specific information of the atoms surrounding the annihilation site can be extracted.
Usually, so-called ratio curves are analyzed, which are obtained (after normalization) by dividing the measured CDB spectra with a reference spectrum (see e.\,g. \cite{Pikart201461}).
The present measurements were performed with the CDB spectrometer \cite{1742-6596-443-1-012071} at the high-intensity positron beam at NEPOMUC \cite{Hugenschmidt2008616,beam}.
The positron implantation energy ranges from $0.5$ to $30$\,keV and spatially resolved measurements can be conducted by scanning the beam across the sample surfaces with a high spatial resolution of $0.3$\,mm for $E>10$\,keV.

\begin{figure}[t]
	\centering
	\includegraphics[width=0.48\textwidth]{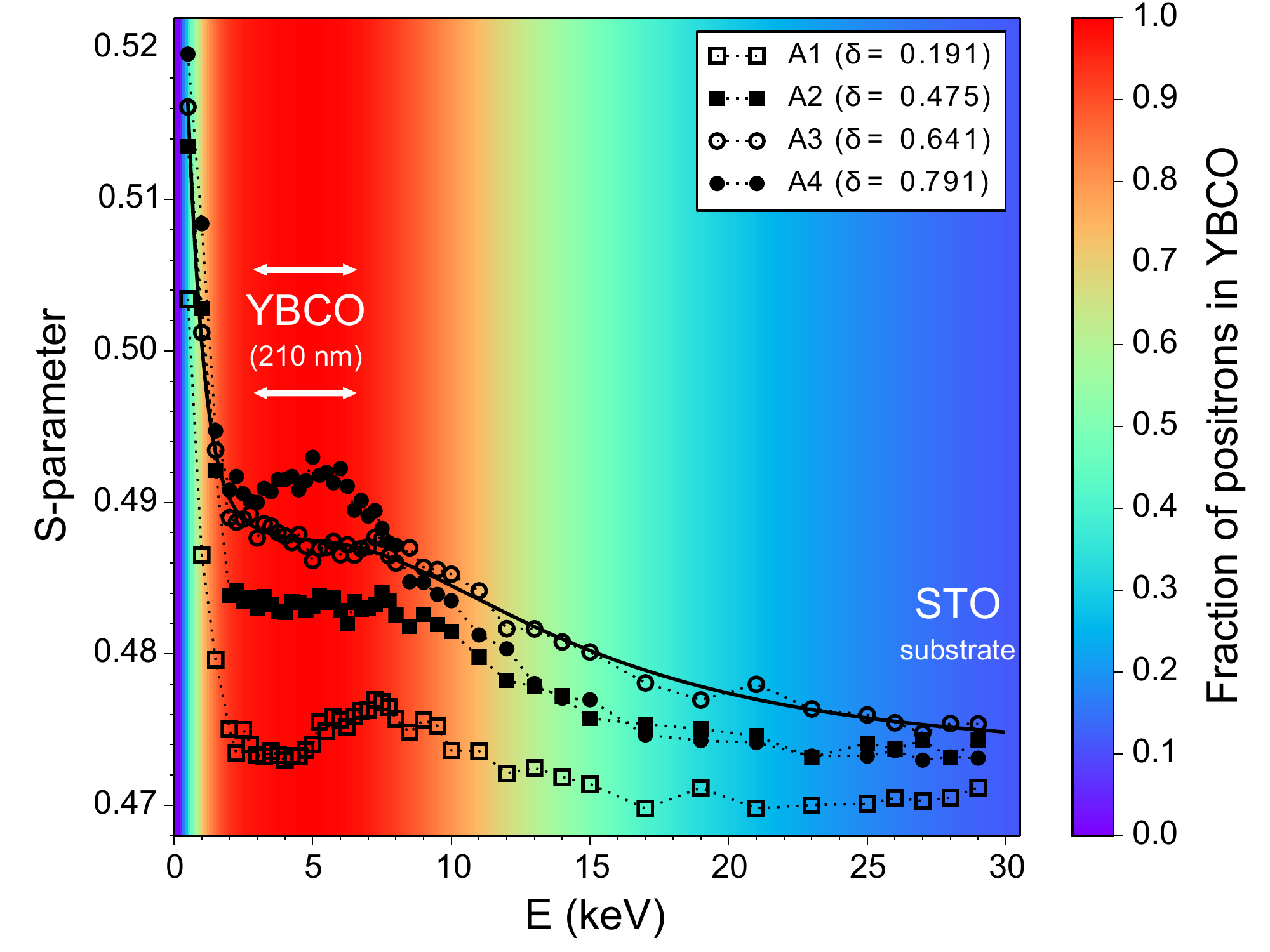}
	\caption{$S(E)$ of the YBa$_2$Cu$_3$O$_{7-\delta}$ thin film samples: The fraction of positrons annihilating in the YBa$_2$Cu$_3$O$_{7-\delta}$ films as function of $E$ is displayed by the color code, which was obtained by the fit shown for sample A3 (solid line).}
	\label{fig:S(E)}
\end{figure}

For all specimens, the $S(E)$ depth profiles were measured (Fig.\,\ref{fig:S(E)}) in order to determine the fraction of positrons annihilating in the YBCO film with the aid of the VEPFIT program \cite{vanVeen1995216}. 
For the least square fits of all $S(E)$ curves, the two-layer YBCO/STO system including the surface was modeled and the respective S-parameters and positron diffusion lengths were fitted.
Exemplary, the fit result for $S(E)$ of sample A3, which yielded the fraction of positrons annihilating in the YBCO film displayed in the color map of Fig.\,\ref{fig:S(E)}, is plotted as solid line. 
The steep increase of $S(E)$ towards the surface ($E < 1.7$\,keV) is explained by the annihilation of positrons after back diffusion to the surface. 
The positron diffusion length $L_+$ was always found to be smaller than 5\,nm, which is extremely short compared to typically 100\,nm for defect-free metallic single crystals (see below).
For high beam energies ($E > 7$\,keV) an increasing fraction of positrons annihilates in the STO substrate and leads to the decrease of $S(E)$. 
The plateau in the range $1.7 < E < 7$\,keV is caused by the positron annihilation predominantly in the YBCO film. 
Therefore, the S-parameter characteristic for each YBCO layer $S_{\mathrm{YBCO}}$ was determined by averaging the S values in the range $3.25 \leq E \leq 6.25$\,keV, where more than $98\,\%$ of the positrons annihilate within the film.
The very flat $S(E)$ dependence between 2\, and 8\,keV of specimen A2 is explained by a high homogeneity in depth of the YBCO film.
In case of sample A1, where the difference between $S_{\mathrm{YBCO}}$ and the substrate S-parameter nearly vanishes, the slight increase 
for $4.5 < E < 7.5$\,keV probably arises from the lattice mismatch ($<2\,\%$) between YBCO and STO leading to a higher defect concentration at the interface. 
A similar but less distinct behavior is also observed in sample A4 between 3 and 6\,keV.

\begin{figure}[t!]
	\centering
	\includegraphics[width=0.48\textwidth]{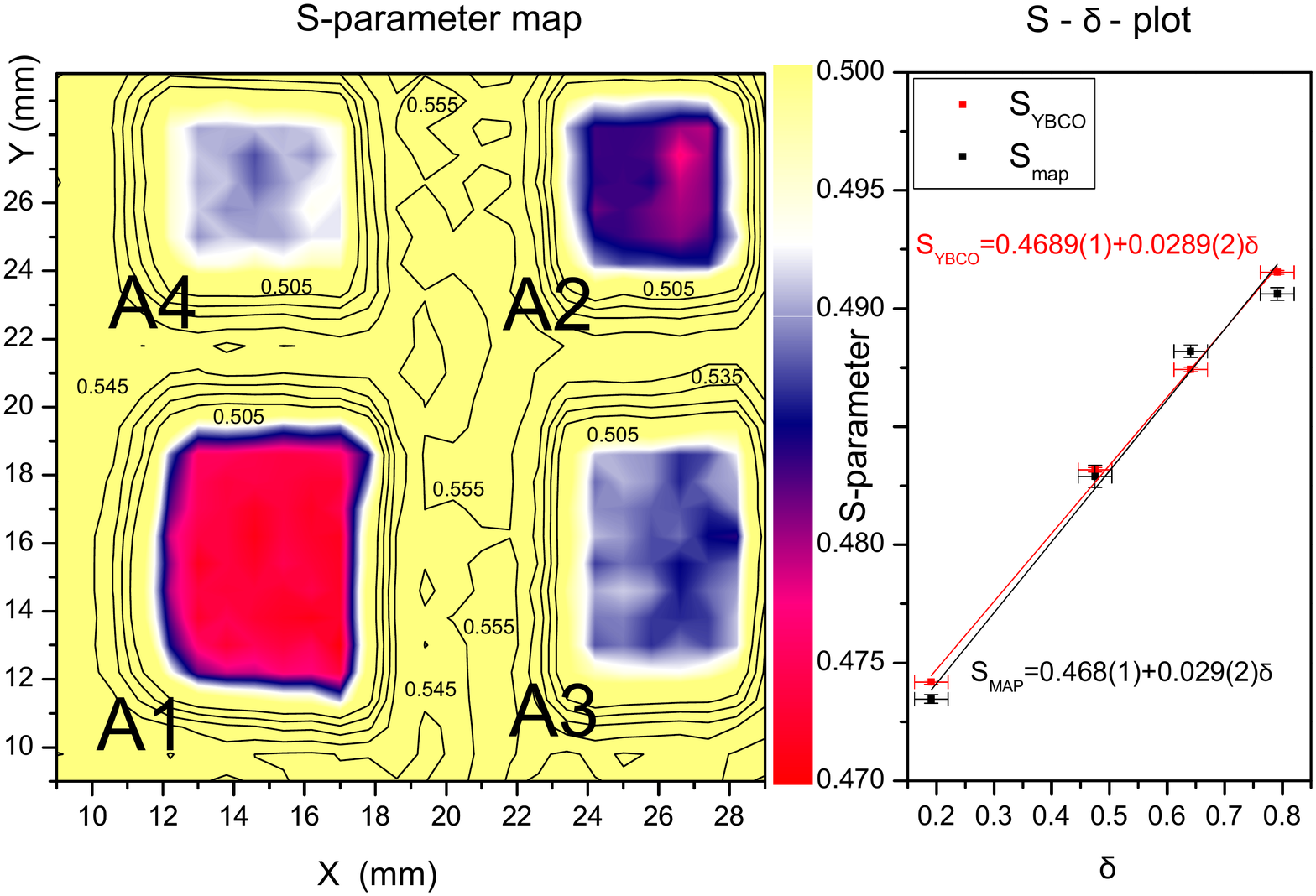}
	\caption{2D S-parameter map of the four YBCO samples obtained by spatially resolved DBS at $E=4$\,keV (left).
		Correlation between S-parameter and $\delta$ determined from XRD (right). The solutions of the linear fits are depicted as solid lines.}
	\label{fig:Sxy}
\end{figure}

In order to probe the structural homogeneity within the plane, spatially resolved DBS was performed by scanning the positron beam over the samples mounted on an Al sample holder (see S(x,y)-map in Fig.\,\ref{fig:Sxy}).
The spatial resolution was 1\,mm at the chosen implantation energy of 4\,keV, where surface and interface effects can be neglected.
For each specimen, a characteristic S-parameter $S_{\mathrm{map}}$ was determined by averaging the S values over the according area in the 2D map.
Only a small difference between $S_{\mathrm{map}}$ and the depth averaged S-parameter $S_{\mathrm{YBCO}}$ was found that is explained by the differently probed depth.
The dependencies of both $S_{\mathrm{YBCO}}$ and $S_{\mathrm{map}}$ on $\delta$ are astonishingly well described by a linear correlation (right plot in Fig.\,\ref{fig:Sxy}).
A similar finding was reported for sintered YBCO bulk samples up to $\delta = 0.6$ \cite{Smedskjaer198856}.

The found correlation $S_{\mathrm{map}}(\delta)$ is used to evaluate the spatial variation $\sigma _{\delta}$ of $\delta$ within the YBCO films.
For this purpose, the spatial scattering of the S-parameter was statistically analyzed for each specimen. 
The standard deviation $\sigma _{\mathrm{map}}$ of the mean value $S_{\mathrm{map}}$ is assumed to comprise both, the \textit{true} spatial variation of the S-parameter $\sigma _{\mathrm{sp}}$ and the statistical error $\sigma _{\mathrm{st}}=1.16\cdot10^{-3}$ of the measured S values.
Consequently, $\sigma _{\mathrm{sp}}$, which is given by $\sigma _{\mathrm{sp}}=\sqrt{\sigma _{\mathrm{map}}^2-\sigma _{\mathrm{st}}^2}$, can be applied to estimate the spatial variation $\sigma _{\delta}$ of $\delta$ with $\sigma _{\delta}=1/0.029(2)\sigma _{\mathrm{sp}}$ using the fit equation shown in Fig.\,\ref{fig:Sxy}. 
It is noteworthy that tempering the YBCO films generally led to a significant increase of the spatial inhomogeneity of $\delta$ (see $\sigma _{\delta}$ values given in Tab.\,\ref{tab:samples}). 
The fluctuation of $\sigma _{\delta} = 0.079(5)$ observed in sample A2 is a factor of four higher than in the non-tempered film A1.
A closer look to the according S(x,y) distribution reveals a region of a lower oxygen deficiency on one side.
Since sample A2 was the only one tempered in an oxygen atmosphere the larger inhomogeneity could be attributed to a more complicated process of oxygen out diffusion \cite{PhysRevB.47.3380}.

\begin{figure}[t!]
	\centering
	\includegraphics[width=0.48\textwidth]{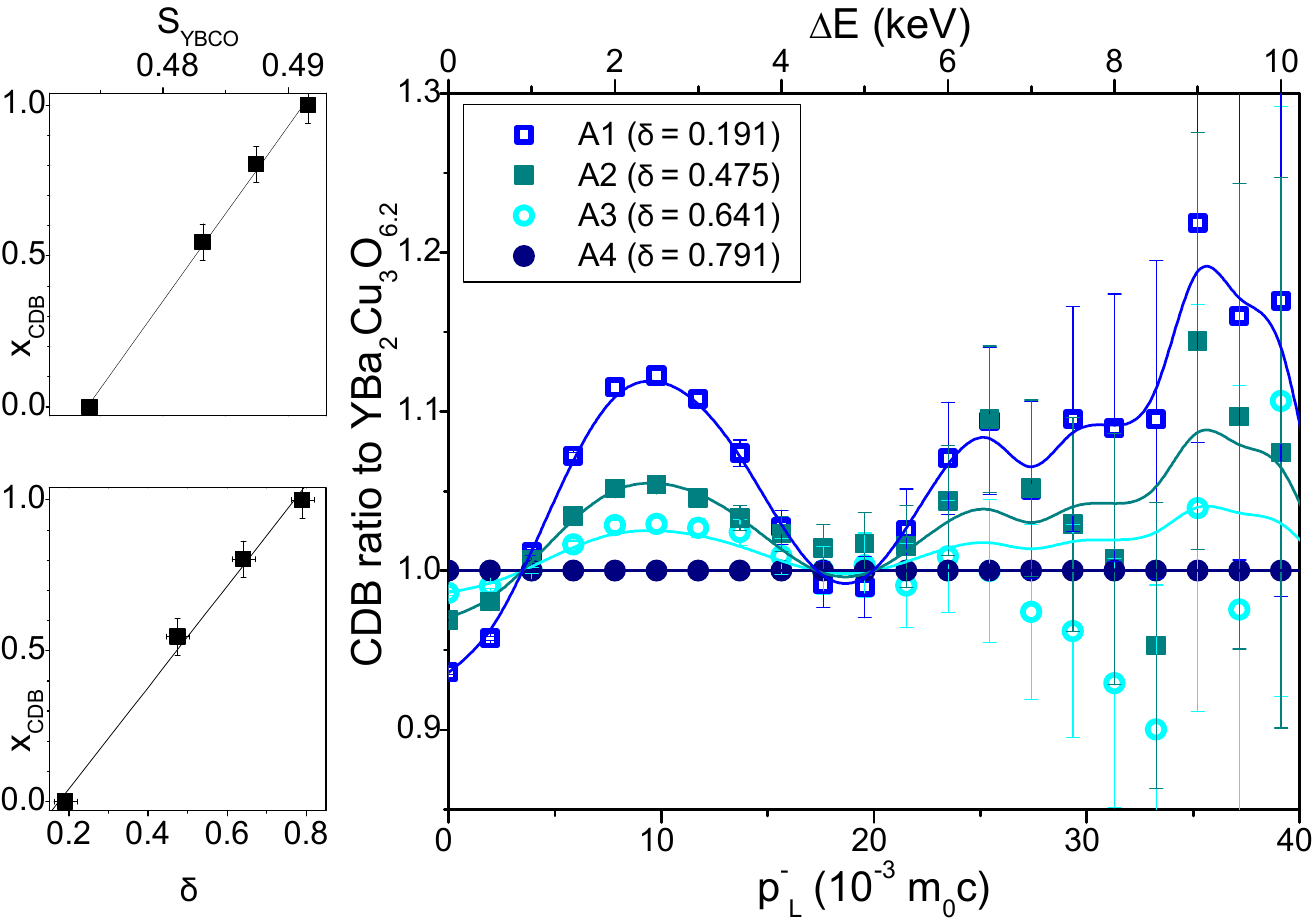}
	\caption{Measured CDBS ratio curves of the YBCO thin film samples (right): The solid lines were obtained by a linear superposition of the spectra A1 and A4 with the fitted weighting factor $x_{\mathrm{CDB}}$. The plots on the left show the linear correlation of $x_{\mathrm{CDB}}$ with $S_{\mathrm{YBCO}}$ and $\delta$, respectively.}
	\label{fig:ratiocurvesA}
\end{figure}

In order to analyze the correlation between the DBS results and $\delta$ in more detail, CDB spectra were recorded with an implantation energy of $E=4$\,keV. 
As shown in Fig.\,\ref{fig:ratiocurvesA}, the measured ratio curves with respect to the sample A4 show systematic changes with decreasing oxygen deficiency: (i) weakening for small electron momenta $p_{\mathrm{L}}^-<4\cdot 10^{-3}$\,m$_0$c, (ii) enhancement for $4\cdot 10^{-3}$\,m$_0$c $< p_{\mathrm{L}}^- < 19\cdot 10^{-3}$\,m$_0$c, and hence higher core annihilation probability and (iii) only tiny enhancement of the element-specific signature in the high momentum region $p_{\mathrm{L}}^- > 19\cdot 10^{-3}$\,m$_0$c.
All spectra $I_{\alpha}\left(p_{\mathrm{L}}^-\right)$ ($\alpha=\mathrm{A1},\dots,\mathrm{A4}$) were fitted using a linear superposition with the weighting factor $x_{\mathrm{CDB}}$:

\begin{equation}
	I_{\alpha}\left(p_{\mathrm{L}}^-\right)=\left(1-x_{\mathrm{CDB}}\right)\cdot I_{\mathrm{A1}}\left(p_{\mathrm{L}}^-\right)+x_{\mathrm{CDB}}\cdot I_{\mathrm{A4}}\left(p_{\mathrm{L}}^-\right)
	\label{eqn:CDBratios}
\end{equation}

The fit results (solid lines in Fig.\,\ref{fig:ratiocurvesA}) well describe the spectra, which essentially show the same signature with different amplitude.
This observation is explained by a transition between two different positron states indicating that tempering of the films leads to a continuous transition from YBa$_2$Cu$_3$O$_{7.00}$ to YBa$_2$Cu$_3$O$_{6.00}$.
In addition, there is no evidence for a drastic $\delta$-dependent change in the structural ordering of oxygen atoms or of the chemical surrounding of the positron annihilation site.
The comparison of the weighting factor $x_{\mathrm{CDB}}$ with both $S_{\mathrm{YBCO}}$ and $\delta$ reveals a strong linear dependence (see Fig.\,\ref{fig:ratiocurvesA}, left).
Thus it can be concluded that DBS and CDBS are clearly sensitive to the varying oxygen deficiency $\delta$, and other effects such as trapping in metallic vacancies, if present, play only a minor role.
For deeper understanding of this behavior, the measured CDB spectra are compared to calculated ones.

\begin{figure}[t!]
	\centering
	\includegraphics[width=0.48\textwidth]{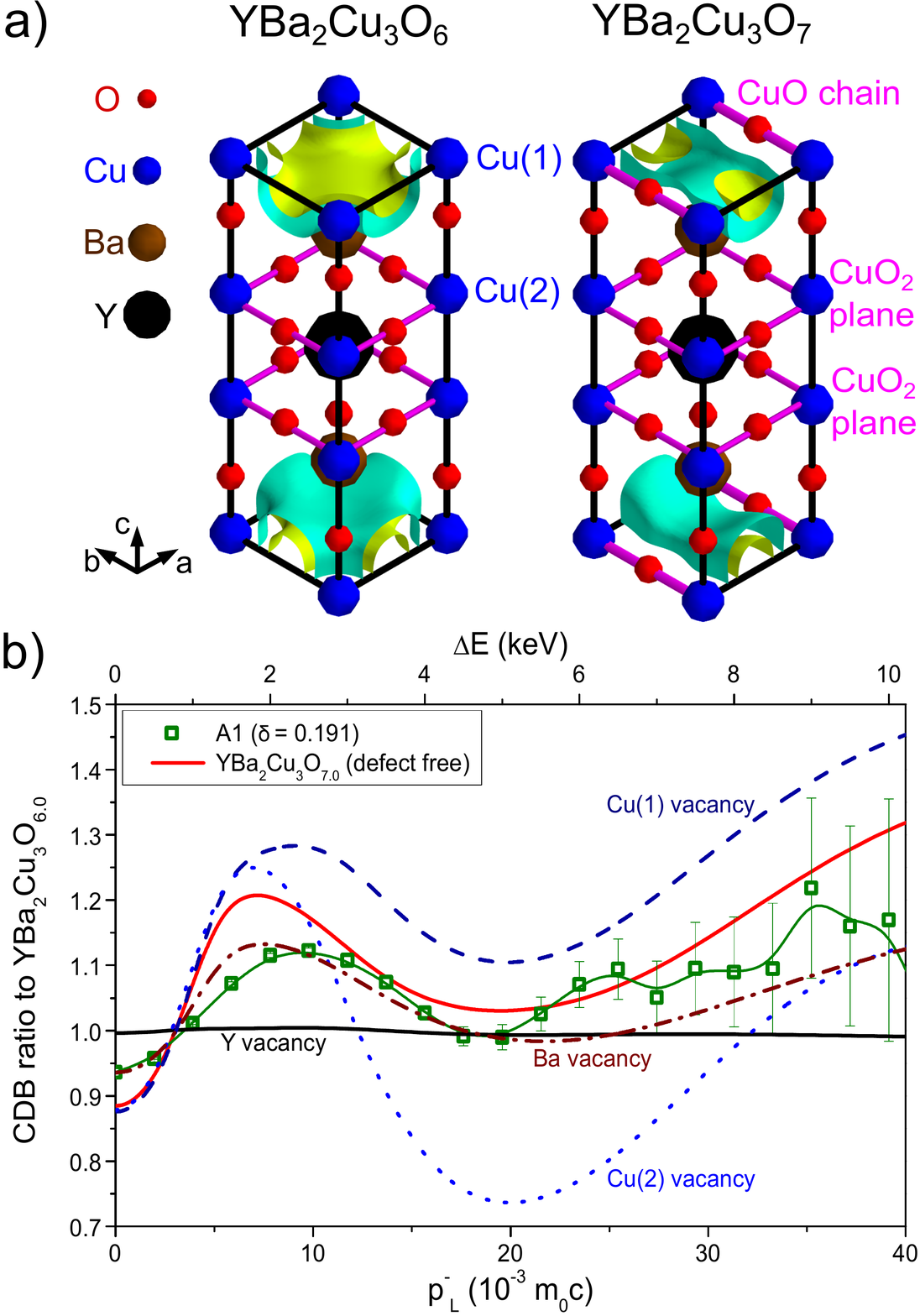}
	\caption{a) Calculated 3D isosurfaces of the positron probability density in YBCO (linear scale of the contours). b) Calculated CDBS ratio curves for defect free YBCO and with different types of metallic vacancies (bold lines). The measured ratio curve of A1/A4 is shown for comparison.}
	\label{fig:ratiocurvesB}
\end{figure}

The CDB spectra were calculated by use of the MIKA Doppler program \cite{PSSB:PSSB200541348}, which describes the positron electron annihilation in a two-component DFT frame in the limit of a vanishing positron density \cite{RevModPhys.66.841} and is based on an atomic superposition method for electron wavefunctions \cite{0305-4608-13-2-009}.
The enhancement of the electron positron correlation is described by a generalized gradient approximation (as proposed by Barbiellini et al. \cite{PhysRevB.51.7341}) with a parametrization based on data by Arponen and Pajanne \cite{a1979343}.
All calculated spectra were folded with a Gaussian in order to mimic the experimental energy resolution as described in \cite{1742-6596-505-1-012025}.

The calculated ratio curve for defect free YBa$_2$Cu$_3$O$_{7.00}$ to YBa$_2$Cu$_3$O$_{6.00}$ (Fig.\,\ref{fig:ratiocurvesB}\,b) exhibits the same features as the measured ones in oxygen rich YBCO thin films. 
The corresponding 3D positron probability densities $\left| \Psi _+ \left( \mathbf{r} \right) \right|^2$ are plotted as isosurfaces in Fig.\,\ref{fig:ratiocurvesB}\,a. 
The calculated 2D delocalized positron state in defect free YBCO is in agreement with previous calculations \cite{PhysRevLett.60.2198,PhysRevB.39.9667,0953-8984-2-6-021,PhysRevB.43.10422,Nieminen19911577,fermischool1,RevModPhys.66.841} for both $\delta$=0 and $\delta$=1.
Within the present calculations, the oxygen deficiency $\delta$ was varied between 0 and 1 under approximation of a tetragonal structure and accounting for the changing lattice constants.
We confirmed that the positron probes the same region in YBCO, i.e. the plane of Cu atoms with more or less oxygen atoms in the Cu-O chains. 
This 2D delocalization of the positron causes a low mobility along the c-axis even in defect free YBCO and hence, could well explain the extremely low positron diffusions length of $L_+ < 5$\,nm observed in the $S(E)$ depth profiles.

Since metallic vacancies can also act as positron trapping sites \cite{Nieminen19911577} their influence on the $\delta$ dependence of the CDB spectra was examined.
For this purpose, the CDB ratio curve of YBa$_2$Cu$_3$O$_{7.00}$ to YBa$_2$Cu$_3$O$_{6.00}$ were calculated including the presence of various vacancy types (Fig.\,\ref{fig:ratiocurvesB}\,b). 
When the positron is trapped in an Y vacancy V$_{Y}$, the CDB spectra for $\delta=0$ and 1 hardly differ and hence lead to a ratio curve equal to unity.
Evidently, this behavior is in contrast to positron trapping in Ba vacancies V$_{Ba}$ or in vacancies at the different Cu sites V$_{Cu(1)}$ and V$_{Cu(2)}$.
The ratio curves for positron annihilation in V$_{Ba}$ or in V$_{Cu(1)}$ exhibit similar features as that for defect-free YBCO: 
Despite being trapped in V$_{Ba}$ the positrons remain sensitive to the changing oxygen content in YBCO, and a positron attracted by a Cu(1) vacancy yields a more enhanced ratio curve above $7\cdot10^{-3}$\,m$_0$c.
A more complex behavior was found for V$_{Cu(2)}$: 
In case of YBa$_2$Cu$_3$O$_{7.00}$, defect trapping clearly leads to the localization of the positron wave function in V$_{Cu(2)}$, whereas V$_{Cu(2)}$ in YBa$_2$Cu$_3$O$_{6.00}$ only leads to a tiny deformation of $\left| \Psi _+ \left( \mathbf{r} \right) \right|^2$, i.\,e. the positron still occupies the CuO plane in a 2D delocalized state.
This effect drastically influences the shape of the obtained ratio curve which exhibits a deep minimum at $20\cdot10^{-3}$\,m$_0$c.
 
The measured $\delta$-dependent ratio curves can be completely explained by annihilation in defect free YBCO but a definite identification of the positron annihilation site is not straight forward.
As revealed by the calculations, the amount of Y vacancies in the YBCO films (if at all) is negligible since positron annihilation in V$_{Y}$ would be independent from $\delta$ and hence would not lead to the strong $\delta$-dependence of the experimental results.
Contrariwise, the presence of Ba and Cu vacancies can not be excluded due to the similarity of the according calculated ratio curves. 
It has to be emphasized, that even in the case of positron trapping in these types of vacancies positrons still sensitively probe the oxygen deficiency in YBCO. 

In this Letter, we demonstrated that (C)DBS using a slow positron beam is a valuable tool for probing YBCO thin films.
The experimental results reveal a strong correlation of the Doppler broadening of the positron annihilation line to the oxygen deficiency $\delta$ determined by XRD.
Backed by calculated CDB spectra, the found linear dependence can be well explained by the positron affinity to the oxygen deficient plane in the defect free YBCO crystal structure.
According to the calculations, surprisingly, a similar dependence can be expected when the positron is trapped in a Ba or Cu vacancy. 
The presence of Y vacancies was found to be unlikely since trapping in this type of vacancy would suppress the positron sensitivity to $\delta$.
Finally, we succeeded to image and to analyse the spatial distribution of the oxygen deficiency quantitatively:
A minimum spatial variation of $\delta$ with $\sigma _{\delta} = 0.019(1)$ was found in the as-deposited film, whereas the $\delta$ fluctuation in tempered films was found to be up to four times larger.

Financial support is gratefully acknowledged within the projects BMBF 05K13WO1 and DFG TRR 80.

%


\end{document}